# Orthonormal RBF wavelet and ridgelet-like series and transforms for high-dimensional problems

W. Chen[*]

Department of Mathematics, City University of Hong Kong, Tat Chee Avenue 83, Kowloon, Hong Kong, China

**Abstract**

This paper developed a systematic strategy establishing RBF on the wavelet analysis, which includes continuous and discrete RBF orthonormal wavelet transforms respectively in terms of singular fundamental solutions and nonsingular general solutions of differential operators. In particular, the harmonic Bessel RBF transforms were presented for high-dimensional data processing. It was also found that the kernel functions of convection-diffusion operator are feasible to construct some stable ridgelet-like RBF transforms. We presented time-space RBF transforms based on non-singular solution and fundamental solution of time-dependent differential operators. The present methodology was further extended to analysis of some known RBFs such as the MQ, Gaussian and pre-wavelet kernel RBFs.

## 1. Introduction

The radial basis function (RBF) was originally introduced in the 1970s to multivariate scattered data approximation and function interpolation. The method has since been applied to neural network, computational geometry, and more recently, numerical partial differential equations (PDE) [1,2]. Its salient merits are independent of dimensionality and geometric complexity, and very mathematically simple to implement. Its complete mathematical theoretical basis is, however, still open. In recent years great effort has been devoted to this task. Some very limited advances are hereto achieved by a sophisticated Hermite-Birkhoff interpolation and native space analysis [2]. There are some essential issues not well answered or even untouched due to great difficulties involved.

Very recently, Chen et al. [3,4] presented kernel RBF based on the intrinsic connection between fundamental solution and nonsingular general solution of differential operator and RBF approximation. The RBF approach is therefore established there on numerical integration of convolution kernel function.

On the other hand, the wavelet is a powerful strategy for mathematical analysis and data processing [5], especially attractive for adaptable handling geometry singularity and localized shock-like solutions due to its inherent multiscale feature combined with a spatial localization. However, applying wavelet to high-dimensional problems still face some challenging issues such as dimensional curse. Fasshauer et al. [6] summarized some wavelets using sphere RBFs. Chen et al. [3,4] proposed super-convergent pre-wavelet kernel RBFs and developed RBFs based on non-singular general solutions of the differential and integral operators. As a logical development of these works, we further derived orthonormal RBF wavelet transforms [4,7].

---

[*] Since Mar. 2001, Dr. Chen will move to Informatics Dept. of Univ. of Oslo, Oslo, Norway.

This paper aims to summarize these new insights on the RBF wavelet transforms. We derived continuous and discrete orthonormal RBF transform analysis for high-dimensional problems and put the RBF on a solid mathematical basis of wavelet analysis. Of them, the harmonic Bessel transform may be broadly useful [7]. Furthermore, we pointed out that the RBF using the non-singular general and fundamental solutions of convection-diffusion operator can be employed as a natural, stable approach to create ridgelet-like RBF transform. To perform time-dependent problems, time-space RBF wavelet analysis was presented which inherently imbeds temporal variable in the generalized RBF. Finally, we use the wavelets to analyze some known RBFs such as the MQ, Gaussian and pre-wavelet kernel RBFs.

## 2. Orthonormal RBF transforms

The goal of this section is to describe the basic theory of discrete orthonormal RBF wavelet transform, and then continuous RBF transform.

Consider a real-value function $f(x)$ on the $n$-dimension spherical domain $\Omega$ of radius R,

$$f(x) = \alpha_0 + \sum_{j=1}^{\infty} \sum_{k=1}^{\infty} \alpha_{jk} \varphi_{nj}(\|x - x_k\|) \quad (1)$$

is its general form of RBF-based wavelet expansion, where $\varphi_{nj}$ represents the wavelet basis function, $\|x - x_k\|$ means Euclidean distance; and $\alpha_{jk}$ are the expansion coefficients. To discrete harmonic analysis, we choose the non-singular general solutions of $n$-dimension Helmholtz equation

$$\varphi_{n0}(r_k) = 1, \qquad \lambda = 0, \quad (2a)$$

$$\varphi_{1j}(r_k) = \frac{1}{2\lambda_j} \sin\left(\frac{\lambda_j r_k}{R}\right), \quad \lambda \neq 0, \quad (2b)$$

$$\varphi_{nj}(r_k) = \left(\frac{\lambda_j}{2\pi r_k}\right)^{(n/2)-1} J_{(n/2)-1}\left(\frac{\lambda_j r_k}{R}\right),$$
$$n \geq 2, \quad \lambda \neq 0 \quad (2c)$$

as wavelet basis function, where $r_k = \|x - x_k\|$, $J_{(n/2)-1}(z)$ is the $(n/2-1)$th order Bessel function of the first kind; and $\lambda_j$ are the zeros of $J_{(n/2)-1}(z)$. The eigenfunctions (2) form a complete orthonormal set of basis functions with weight function $r^{n-1}$, and therefore, the approximation of any piecewise smooth functions using these bases is absolutely uniformly convergence. Such RBF wavelet series is defined as the discrete Bessel transform (DBT) of function $f(x)$, simply called as $B$-transform. It is known that the Bessel function possesses the orthonormal merit, i.e.,

$$\int_0^a z J_v(\eta z) J_v(\mu z) dz = \begin{cases} 0 & \eta \neq \mu \\ C & \eta = \mu \end{cases}, \quad (3)$$

where $C = [J_{v+1}(\eta)^2]/2$ under $a=1$, $\eta=\mu$, and $J_v(\eta) = J_v(\mu) = 0$. We can thereby determine the expansion coefficients for problems of more than two dimensions by

$$\alpha_0 = \frac{1}{R^n} \int_{\Omega_R} f(\zeta) d\Omega_\zeta, \quad (4a)$$

$$\alpha_{jk} = \frac{2}{R^{n+1} J_{n/2}(\lambda_j)^2} \left(\frac{\lambda_j}{2\pi}\right)^{1-n/2} \int_{\Omega_\zeta} r_{k\zeta}^{n/2} f(\zeta)$$
$$J_{(n/2)-1}\left(\frac{\lambda_j r_{k\zeta}}{R}\right) d\Omega_\zeta, \quad j=1,2,\ldots, \quad k=1,2,\ldots,$$
$$(4b)$$

where $r_{k\zeta} = \|x_k - x_\zeta\|$, $R$ is the radius of the spherical domain centering node $k$. It is obvious that the DBT is an orthonormal wavelet expansion in terms of one-dimensional Euclidean distance variable.

Substituting Eqs. (4a,b) into formula (1) produces

$$f(x) = \frac{1}{R^n} \int_{\Omega_R} f(\zeta) d\Omega_\zeta + \sum_{j=1}^{\infty} \sum_{k=1}^{\infty} \frac{2}{R^{n+1} J_{n/2}(\lambda_j)^2}$$
$$\int_{\Omega_R} r_{k\zeta}^{n/2} f(\zeta) J_{(n/2)-1}\left(\frac{\lambda_j r_{k\zeta}}{R}\right) d\Omega_\zeta \, r_{xk}^{1-n/2} J_{(n/2)-1}\left(\frac{\lambda_j r_{xk}}{R}\right)$$
$$(5)$$

Furthermore, in terms of continuous wavelet transform theory and assuming $f(x)$ absolutely integrable in $\Omega_\infty$, we have

$$F(\lambda,\xi) = \int_{\Omega_\infty} r_{\xi\zeta}^{n-1} f(\zeta) \varphi_n(\lambda r_{\xi\zeta}) d\Omega_\zeta, \quad (6)$$

$$f(x) = \frac{1}{C_{\varphi_n}} \int_0^{+\infty} \int_{\Omega_\infty} F(\lambda,\xi) \varphi_n(\lambda r_{x\xi}) \lambda^m d\Omega_\xi d\lambda, \quad (7)$$

where $C_{\varphi_n}$ and $m$ can be derived by using the continuous wavelet theory [5]. Eqs. (6) and (7) are defined as continuous Bessel integral or transform of function $f(x)$.

It is noted that the locality of the common wavelet expression is implicitly built in RBF distance variable and these RBF wavelets are inherently symmetric. More importantly, this RBF high-dimensional wavelet construction is based on God-blessing natural orthonormal kernel RBF rather than on the traditional tensor product approach embattled by the curse of dimensionality [8].

## 3. Bi-orthogonal RBF transforms

To construct bi-orthogonal wavelets, we need to find the dual wavelet basis functions. It is known that the two general solutions of the $n$-dimension Helmholtz operator are

$$\varphi_n(\lambda r_k) = \left(\frac{\lambda}{2\pi_k r_k}\right)^{(n/2)-1} H^{(1)}_{(n/2)-1}(\lambda r_k), \; n \geq 2, \quad (8)$$

and

$$\phi_n(\lambda r_k) = \left(\frac{\lambda}{2\pi_k r_k}\right)^{(n/2)-1} H^{(2)}_{(n/2)-1}(\lambda r_k), \; n \geq 2, \quad (9)$$

where $H^{(1)}$ and $H^{(2)}$ respectively denote the Hankel function of the first kind and the Hankel function of the second kind, namely,

$$H^{(1)}_{(n/2)-1}(\lambda r_k) = J_{(n/2)-1}(\lambda r_k) + i Y_{(n/2)-1}(\lambda r_k), (10)$$

$$H^{(2)}_{(n/2)-1}(\lambda r_k) = J_{(n/2)-1}(\lambda r_k) - i Y_{(n/2)-1}(\lambda r_k). (11)$$

$Y_{(n/2)-1}(r)$ is the Bessel function of second kind. Note that these two Helmholtz general solutions are essentially orthogonal. It is obvious

$$H^{(1)}_{(n/2)-1}(\lambda r_k) = \overline{H^{(2)}_{(n/2)-1}(\lambda r_k)}. \quad (12)$$

In addition, we have

$$H^{(1)}_{(n/2)-1}(\lambda r_k) = \frac{2}{i\pi} K_{(n/2)-1}(-i\lambda r_k), \quad (13)$$

which $K$ denotes the modified Bessel function of the second kind. In this case, it is noted that the functions $H^{(1)}$, $H^{(2)}$, $Y$, and $K$ encounter singularity at zero. However, their integrals always exist.

One can easily conclude that the proper harmonic wavelet should be

$$g_n(\lambda r_k) = \frac{1}{2\pi} \left(\frac{-i\lambda}{2\pi r_k}\right)^{(n/2)-1} K_{(n/2)-1}(-i\lambda r_k). \quad (14)$$

The corresponding dual wavelet basis function is its conjugate function $\overline{g_n(\lambda r_k)}$. The RBF bi-orthogonal wavelet transform is thus established by

$$F(\lambda,\xi) = \int_{\Omega_\infty} r_{\xi\zeta}^{n-1} f(\zeta) \overline{g_n(\lambda r_{\xi\zeta})} d\Omega_\zeta, \quad (15)$$

and

$$f(x) = C_g^{-1} \int_{-\infty}^{+\infty} \int_{\Omega_\infty} F(\lambda,\xi) g_n(\lambda r_{x\xi}) \lambda^m d\Omega_\xi d\lambda, \quad (16)$$

where $C_g$ and $m$ are decided by the normal approach in continuous wavelet transform. We call the above integral transform involving the modified Bessel function $K$ as the $K$-transform.

The $B$- and $K$-transforms carry some nice features of Fourier and wavelet transforms. In particular, they hold

$$B[\Delta f(x)] = -\lambda^2 B[f(x)], \quad (17)$$

$$K[\Delta f(x)] = -\lambda^2 K[f(x)], \quad (18)$$

where $B[\;]$ and $K[\;]$ respectively denote the $B$-

and *K*-transforms. An immediate application is to solve higher-dimensional PDE systems.

In fact, $\sin(\lambda r)$, $\cos(\lambda r)$ and $e^{\pm i\lambda r}$ may directly be employed as the wavelet basis function in RBF transforms. However, those functions do not involve dimensional affect and do not satisfy formulas (17) and (18).

## 4. Ridgelet-like RBF transforms

As pointed out by Candµes and Donoho [9], the ridge functions have been in recent years widely applied in approximation theory and statistics. An superposition of ridge functions may converge to an underlying function *f(x)* of high dimensions at good rates.

$$\hat{f}_N(x) = \sum_{k=1}^{N} \alpha_k \sigma(\mu_k \bullet x - b_k), \quad (19)$$

where $\alpha$, $\mu$, and *b* respectively represent the scale, direction, and locations; $\sigma$ is the ridge function. The wavelet is found to be a powerful tool constructing stable ridgelet approximation. Some robust orthonormal wavelet-based ridge functions

$$\psi_{\alpha,\mu,b}(x) = \alpha^{1/2}\psi(\alpha\mu x - b) \quad (20)$$

are thereby created [9]. This section attempts to make some ridgelet-like RBF transforms by using orthonormal kernel functions.

In preceding analysis, we actually use the fundamental solutions and non-singular general solutions of Helmholtz and modified Helmholtz operators. In fact, there exists more flexibility using various orthogonal convolution kernel functions of linear and nonlinear differential and integral operators to construct RBF wavelets. In the case of the ridgelets, let us consider the *n*-dimension convection-diffusion equation

$$D\nabla^2 u + v \bullet \nabla u - ku = 0. \quad (21)$$

where *v* denotes velocity vector, *D* is the diffusivity coefficient, and *k* represents the reaction coefficient. Its non-singular general solution is

$$u_n(r) = \frac{1}{2\pi}\left(\frac{\mu}{2\pi r}\right)^{(n/2)-1} e^{-\frac{v \cdot r}{2D}} I_{(n/2)-1}(\mu r), \quad (22)$$

and fundamental solution

$$u_n^*(r) = \frac{1}{2\pi}\left(\frac{\mu}{2\pi r}\right)^{(n/2)-1} e^{-\frac{v \cdot r}{2D}} K_{(n/2)-1}(\mu r), \quad (23)$$

where

$$\mu = \left[\left(\frac{|v|}{2D}\right)^2 + \frac{k}{D}\right]^{\frac{1}{2}}. \quad (24)$$

The smooth function (22) can be used in orthonormal discrete and continuous RBF wavelet transforms, while the corresponding bi-orthogonal RBF wavelets are

$$g_n(\mu r) = \frac{1}{2\pi}\left(\frac{\mu}{2\pi r}\right)^{(n/2)-1} e^{-\frac{v \cdot r}{2D}} N_{(n/2)-1}(\mu r), \quad (25)$$

and its conjugate function, where

$$N_{(n/2)-1}(\mu r) = I_{(n/2)-1}(\mu r) + iK_{(n/2)-1}(\mu r). \quad (26)$$

Note that velocity vector *v* is much like the role of direction vector $\mu$ in ridge function (19). Therefore, we can construct ridgelet-like RBF wavelet expansions which naturally combines direction gradient affect using RBFs (22) and (25). It is worth pointing out that the convection-diffusion problems are closely related to phenomena of shock and solutions with localized great gradient variations. Therefore, such RBF transform may be applicable to edge data processing.

It is very interesting to observe that these convection-diffusion ridgelet-like RBFs have very similar behavior of the Gaussian. However, there is no existing method for constructively creating a stable and effective Gaussian approximation, while these RBF wavelet transforms play operationally the same role. We can use these transforms combined with various neural network and least squares techniques to estimate characteristic parameters of very high-dimension problem. In the convection-diffusion wavelets, $D$, $v$ and $k$ are desirable recognized parameters.

Furthermore, $n$ in the foregoing analysis is set as a positive integer index of dimensionality. It is imaginable that $n$ can be generalized to a real number corresponding to a fractal dimensionality, namely, fractal RBF discrete and continuous wavelet transforms, which may be significant to irregular data processing.

## 5. Time-space RBF transforms

Another important issue we will deal with is time-dependent system. The present author has recently proposed time-space RBF based on the fundamental solutions and nonsingular general solutions of time-dependent wave, diffusion, and convection-diffusion operators [3,4]. Namely, we can generalize the definition of RBF to intrinsically imbed temporal variable. The non-singular general solutions for $n$-dimensional heat and diffusion problems are

$$\varphi_1(\lambda r_k, \alpha \Delta t_k) = H(\Delta t_k) e^{-a^2 \lambda^2 \Delta t_k} \frac{\sin(\lambda r_k)}{2\lambda}, \quad (27a)$$

$$\varphi_n(\lambda r_k, \alpha \Delta t_k) = H(\Delta t_k) e^{-a^2 \lambda^2 \Delta t_k} \left(\frac{\lambda}{2\pi r_k}\right)^{(n/2)-1} J_{(n/2)-1}(\lambda r_k), \quad n \geq 2 \quad (27b)$$

and for wave problems

$$\phi_1(\lambda r_k, c\Delta t_k) = \left[\alpha_k \cos(c\lambda \Delta t_k) + \frac{\beta_k}{c\lambda} \sin(c\lambda \Delta t_k)\right] \frac{\sin(\lambda r_k)}{2\lambda} H(\Delta t_k) H(c\Delta t_k - r_k) \quad (28a)$$

$$\phi_n(\lambda r_k, c\Delta t_k) = \left[\alpha_k \cos(c\lambda \Delta t_k) + \frac{\beta_k}{c\lambda} \sin(c\lambda \Delta t_k)\right] \left(\frac{\lambda}{2\pi r_k}\right)^{(n/2)-1} J_{(n/2)-1}(\lambda r_k) H(\Delta t_k) H(\lambda \Delta t_k - r_k), \quad n \geq 2, \quad (28b)$$

where $\Delta t_k = t - t_k$, $n$ denotes dimensionality, $a$ and $c$ are respectively physical coefficients of diffusion and wave problems, $\lambda \neq 0$, $H$ represents Heaviside function. The above RBFs can be used to construct the discrete and continuous RBF wavelet transform. For instance, the continuous RBF transform for problems having transient diffusion backgrounds are built by

$$F(\lambda, \xi) = \int_0^{+\infty} \int_{\Omega_\infty} r_{\xi\zeta}^{n-1} f(\zeta, t) \varphi_n(\lambda r_{\xi\zeta}) d\Omega_\zeta dt \quad (29)$$

and

$$f(x,t) = \frac{1}{C_{\varphi_n}} \int_0^{+\infty} \alpha^2 \lambda^2 \int_0^{+\infty} e^{-a^2\lambda^2(t-\tau)} H(t-\tau) d\tau \int_{\Omega_\infty} F(\lambda, \xi) \varphi_n(\lambda r_{x\xi}) \lambda^m d\Omega_\xi d\lambda \quad (30)$$

where $C_\varphi$, $\varphi_n$, and $m$ are the same as those in Eq. (7). Similarly, we can derive the RBF transform using functions (28a, b) for wave problems such as seismic data processing.

The bi-orthogonal transform for dynamic diffusion and wave problems can be achieved respectively by using

$$\varphi_n(\lambda r_k, \alpha \Delta t_k) = H(\Delta t_k) e^{-a^2 \lambda^2 \Delta t_k} \left(\frac{\lambda}{2\pi r_k}\right)^{(n/2)-1} K_{(n/2)-1}(-i\lambda r_k), \quad n \geq 2, \quad (31)$$

$$\phi_n(\lambda r_k, c\Delta t_k) = \left[\alpha_k \cos(c\lambda \Delta t_k) + \frac{\beta_k}{c\lambda} \sin(c\lambda \Delta t_k)\right] \left(\frac{\lambda}{2\pi r_k}\right)^{(n/2)-1} K_{(n/2)-1}(-i\lambda r_k) H(\Delta t_k) H(\lambda \Delta t_k - r_k),$$
$$n \geq 2 \quad (32)$$

and their conjugate functions. In addition, the corresponding time-dependent fundamental solutions [4] may be also feasible for the same task. It is noted that transient problems can be handled in the same way as for steady problems with the heat and wave potentials concepts.

## 6. Wavelet analysis of known RBFs

In terms of wavelet analysis, the known MQ can be written as

$$f(x) = \alpha_0 + \sum_{l=1}^{n} \alpha_l x^{(l)} + \sum_{j=1}^{N} \sum_{k=1}^{M} \beta_{jk} \sqrt{r_k^2 + c_j^2}. \quad (33)$$

The shape parameters $c_j$ are here interpreted as the scales of the RBF interpolation. The MQ RBF is not orthonormal. Therefore, we can not evaluate its expansion coefficients explicitly as in the foregoing cases. $\sqrt{r_k^2 + c^2}$ in which way the MQ is usually applied is seen as the simplified version of RBF expansion (33). The normal Gaussian RBF is a similar case but without plus polynomial terms.

On the other hand, Chen et al. [3] found that if we replace distance variable $r_k$ by $\sqrt{r_k^2 + c^2}$ in the kernel RBF, we got the pre-wavelet RBF with the exponential accuracy, where $c$ is called as scale (dilution) parameter. Numerical experiments with pre-wavelet TPS $(r_k^2 + c^2)\ln\sqrt{r_k^2 + c^2}$ manifested clearly spectral convergence like the MQ. Obviously, such form of pre-wavelet TPS should be understood as the simplified RBF wavelet expansion.